\documentclass{amia}
\usepackage{graphicx}
\usepackage[labelfont=bf]{caption}
\usepackage[superscript,nomove]{cite}
\usepackage{scrextend}
\usepackage{color,soul}
\usepackage[table,xcdraw]{xcolor}
\usepackage{hyperref}
\usepackage{multirow}
\hypersetup{
    colorlinks=true,
    linkcolor=blue,
    filecolor=magenta,      
    urlcolor=blue,
}

\begin{document}

\title{Intelligent Word Embeddings of Free-Text Radiology Reports}

\author{Imon Banerjee, Ph.D$^{1}$, Sriraman Madhavan, B.E.$^{1}$, Roger Eric Goldman, M.D., Ph.D.$^{1}$, Daniel L. Rubin, M.D.$^{1}$}

\institutes{
    $^{1}$Department of Radiology, Stanford University School of Medicine, Stanford, USA 
\\
}

\maketitle

\noindent{\bf Abstract}

\textit{Radiology reports are a rich resource for advancing deep learning applications in medicine by leveraging the large volume of data continuously being updated, integrated, and shared. However, there are significant challenges as well, largely due to the ambiguity and subtlety of natural language. We propose a hybrid strategy that combines semantic-dictionary mapping and word2vec modeling for creating dense vector embeddings of free-text radiology reports. Our method leverages the benefits of both semantic-dictionary mapping as well as unsupervised learning. Using the vector representation, we automatically classify the radiology reports into three classes denoting confidence in the diagnosis of intracranial hemorrhage by the interpreting radiologist. We performed experiments with varying hyperparameter settings of the word embeddings and a range of different classifiers. Best performance achieved was a weighted precision of 88\% and weighted recall of 90\%. Our work offers the potential to leverage unstructured electronic health record data by allowing direct analysis of narrative clinical notes.}

\section{Introduction}
The Picture Archiving and Communication Systems (PACS) stores a wealth of unrealized potential data for the application of deep learning algorithms that require a substantial amount of data to reduce the risk of overfitting. Semantic labeling of data becomes a prerequisite to such applications. Each PACS database serving a major medical center contains millions of imaging studies ``labeled'' in the form of unstructured free text of the radiology report by the radiologists, physicians trained in medical image interpretation. However, the unstructured free text cannot be directly interpreted by a machine due to the ambiguity and subtlety of natural language and variations among different radiologists and healthcare organizations. Lack of labeled data creates data “bottleneck” for the application of deep learning methods to medical imaging \cite{WangLSKNYS16}. 

In recent years, there is movement towards structured reporting in radiology with the use of standardized terminology~\cite{kahn2009toward}. Yet, the majority of radiology reports remain unstructured and use free-form language. To effectively ``mine" these large free-text data sets for hypotheses testing, a robust strategy for extracting the necessary information is needed. Methods for structuring and labeling the radiology reports in the PACS may serve to unlock this rich source of medical data. 

Extracting insights from free-text radiology reports has been explored in numerous ways. Nguyen et al.\cite{nguyen2014supervised} combined traditional supervised learning methods with Active Learning for classification of imaging examinations into reportable and non-reportable cancer cases. Dublin et al.\cite{dublin2013natural} and Elkin et al.\cite{elkin2008nlp} explored sentence-level medical language analyzers and SNOMED CT-based semantic rules respectively, to identify pneumonia cases from free-text radiological reports. Huang et al.\cite{huang2007novel} introduced a hybrid approach that combines semantic parsing and regular expression matching for automated negation detection in clinical radiology reports.

In recent years, the word2vec model introduced by Mikolov et al. \cite{mikolov2013efficient, mikolov2013distributed} has gained interest in providing semantic word embeddings. One of the biggest problems with word2vec is the inability to handle unknown or out-of-vocabulary~(OOV) words and morphologically similar words. The challenge is exacerbated in domains, such as radiology, where synonyms and related words can be used depending on the preferred style of radiologist, and a word may only have been used infrequently in a large corpus. If the word2vec model has not encountered a particular word before, it will be forced to use a random vector, which is generally far from its ideal representation. Thus, we explore how the word2vec model can be combined with the radiology domain-specific semantic mappings in order to create a legitimate vector representation of free-text radiology reports. The application we have explored is the classification of reports by confidence in the diagnosis of intracranial hemorrhage by the interpreting radiologist. 

\clearpage
Our two core contributions are:
\begin{enumerate}
    \item We proposed a hybrid technique for a dense vector representation of individual words of the radiology reports by analyzing 10,000 radiology reports associated with computed tomography (CT) Head imaging studies. \\
    \textit{(Word embeddings are publicly released in: \href{https://github.com/imonban/RadiologyReportEmbedding} {https://github.com/imonban/RadiologyReportEmbedding})}
    \item Using our methods, we automatically categorized radiology reports according to the likelihood of intracranial hemorrhage.  
\end{enumerate}

We derived the word embeddings from large unannotated corpora that were retrieved from PACS (10,000 reports), and the classifiers were trained on a small subset of annotated reports (1,188). The proposed embedding produced high accuracy (88\% weighted precision and 90\% recall) for automatic multi-class (low, intermediate, high) categorization of free-text radiology reports despite the fact that the reports were generated by numerous radiologists of differing clinical training and experience. We also explored the visualization of vectors in low dimensional space while retaining the local structure of the high-dimensional vectors, to investigate the legitimacy of the semantic and syntactic information of words and documents. In the following sections, we detail the methodology (Sec.~\ref{sec:methodology}), present the results (Sec.~\ref{sec:result}) and finally conclude by mentioning future directions (Sec.~\ref{sec:conclusion}).

\section{Methodology \label{sec:methodology}}

\begin{figure}[H]
\centering
\includegraphics[scale=0.4]{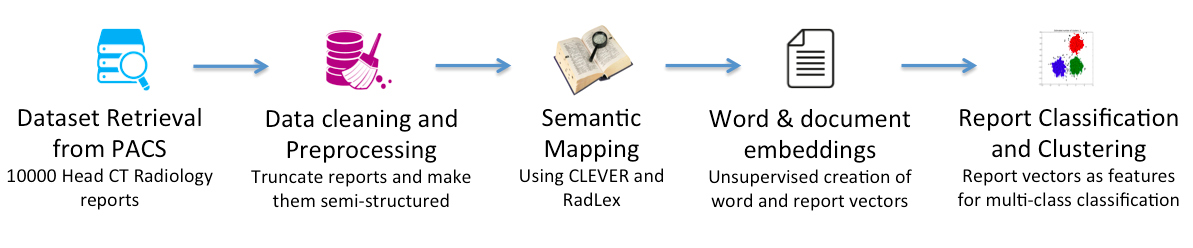}
\caption{Components of the proposed framework}
\label{fig:methodology}
\end{figure}

Figure~\ref{fig:methodology} shows the proposed research framework that comprises five components: Dataset retrieval from PACS, Data Cleaning \& Preprocessing, Semantic-dictionary mapping, Word and Report Embedding, and Classification. In the following subsections, we describe each component.

\subsection{Dataset \label{Sec:dataset}}
The dataset consists of the radiology reports associated with all computed tomography (CT) studies of the head located in the PACS database serving of our adult and pediatric hospitals and all affiliated outpatient centers for the year of 2015. Through an internal custom search engine, candidate studies were identified on the PACS server based on imaging exam code. The included study codes captured all CT Head, CT Angiogram Head, and CT Head Perfusion studies. A total of 10,000 radiology reports were identified for this study. In order to provide a gold standard reference for the vector-space embedding algorithm, a subset of 1,188 of the radiologic reports were labeled independently by two radiologists. For each report, the radiologists read the previous interpretation and then graded the confidence of the interpreting physician with respect to the diagnosis of intracranial hemorrhage. For each study, a numeric label was provided on a scale ranging from 1 to 5 with labels as follows: 1) No intracranial hemorrhage; 2) Diagnosis of intracranial hemorrhage unlikely, though cannot be completely excluded; 3) Diagnosis of intracranial hemorrhage possible; 4) Diagnosis of intracranial hemorrhage probable, but not definitive; 5) Definite intracranial hemorrhage. These labels were chosen to reflect heuristics employed by radiologists and treating physicians to interpret the spectrum of information produced by the imaging study.

\subsection{Data Cleaning \& Preprocessing}

All 10,000 radiology reports were transformed through a series of pre-processing steps to truncate the free-text radiology reports and to focus only on the significant concepts, which would enhance the semantic quality of the resulting word embeddings. We developed a python-based text processor - \textit{Report Condenser}, that executes the pre-processing steps sequentially. First, it extracted the \textit{`Findings'} and \textit{`Impressions'} sections from each report that summarizes the CT image interpretation outcome, since our final objective was to classify the reports based on radiological findings.

In the next pre-processing stage, the Report Condenser cleansed the texts by normalizing the texts to lowercase letters and removing words of following types: general stop words, words with very low frequency ($\textless 50$), unwanted terms and phrases (e.g. medicolegal phrases - \textit{``I have personally reviewed the images for this examination and agreed with the report transcribed above."}, headers - \textit{`FINDINGS', `IMPRESSION', `Additional comment'}). These words usually appear either in all the reports or in a very few reports, thus of little or no value in document classification. We used the NLTK library \cite{bird2006nltk} for determining a stop-word list and discarded them during indexing. Examples of the stop-words are: \textit{a, an, are,...,be, by,...,has, he,...,etc}. The Report Condenser also discarded datestamps, timestamps, the radiologist details (e.g. names, contacts) and other recurring phrases in reports. Removal of these terms significantly reduced the number of words that the system had to handle.

Following the removal steps, Report Condenser searched the updated corpus to identify frequently appearing pairs of words based on pre-defined threshold value of occurrence ($>500$) and concatenated them into a single word to preserve useful semantic units for further processing. Some examples of the concatenated words are: \textit{`midline shift' $\rightarrow$ `midline\_shift',  `mass effect'  $\rightarrow$  `mass\_effect',
`focal abnormality' $\rightarrow$ `focal\_abnormality'}.

In the next step, Report Condenser identified and encoded negation dependencies that appear in the radiology reports via simple string pattern matching. For example, in the phrase \textit{`No acute hemorrhage, infarction, or mass'}, negation is applied to `acute hemorrhage', `infarction' as well as `mass'. Therefore, the Report Condenser encodes the negation dependency as: \textit{`No\_acute\_hemorrhage', `No\_infarction', `No\_mass'}. Such phrases were identified automatically by analyzing the whole corpus and transformed accordingly.

\subsection{Semantic-dictionary mapping \label{sec:dicmapping}}
The main idea of the Semantic-dictionary mapping is to use a lexical scanner that
recognizes corpus terms which share a common root or stem with pre-defined terminology, and map them to controlled terms. In contrast with traditional NLP approaches, this step does not need any sentence parsing, noun-phrase identification, or co-reference
resolution. We used dictionary style string matching where we directly search and replace terms, by referring to the dictionary. We implemented a lexical scanner in python which can handle 1 kilobyte of text per millisecond. On average, the size of each radiology report after cleaning was 1 kilobyte and our scanner took less than 10 seconds to complete the whole mapping process for 10,000 radiology reports. We applied the following two-stage process. 

1. \textit{Common terms mapping}: First, we used the more general publicly available \href{https://github.com/stamang/CLEVER/blob/master/res/dicts/base/clever_base_terminology.txt}{CLEVER terminology}~\cite{jung2013automated} to replace common analogies/synonyms for creating more semantically structured texts. We focused on the terms that describe family, progress, risk, negation, and punctuations, and normalized them using the formal terms derived from the terminology.

For instance, \{\textit{`mother', `brother', `wife' ..} \}$\,\to\,$`FAMILY', \{\textit{`no', `absent', `adequate to rule her out' .. }\}$\,\to\,$`NEGEX', \{\textit{`suspicion', `probable', `possible'} \}$\,\to\,$`RISK', \{\textit{`increase', `invasive', `diffuse', .. }\}$\,\to\,$`QUAL'.

2. \textit{Domain-specific dictionary mapping}: For this case-study,  
we used the domain-specific RadLex ontology~\cite{mejino2008fma} for mapping the variations of radiological terms that are related to hemorrhage, to a controlled terminology.  We created an ontology crawler using SPARQL that grabs the sub-classes and synonyms of the domain-specific terms from Radlex, and creates a focused dictionary for \textit{``Intracranial hemorrhage"} radiology reports. Using the dictionary all the equivalent terms of hemorrhage are formalized in the corpus as: \textit{\{`apoplexy', `contusion', `hematoma', ... \}$\,\to\,$'hemorrhage'}.

\begin{figure}[H]
\centering
\includegraphics[width=\linewidth]{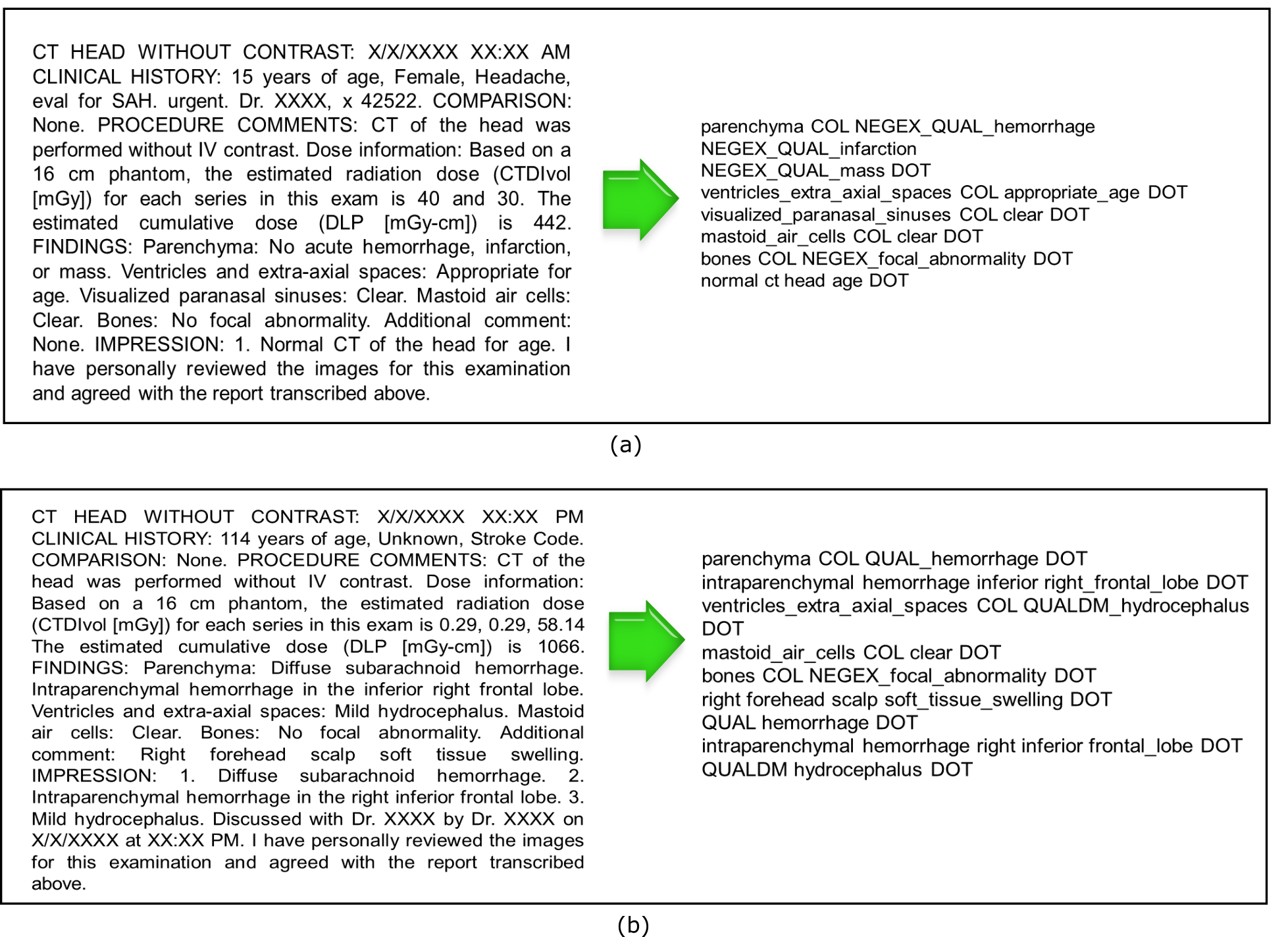}
\caption{Examples of preprocessing and semantic-dictionary mapping - on the left FINDINGS and IMPRESSION sections of the original reports and on the right processed reports of (a) low and (b) high likelihood of intracranial hemorrhage. (Names and dates have been redacted to preserve anonymity)}
\label{fig:preprocessing}
\end{figure}
In Figure~\ref{fig:preprocessing}, we present the outcome of preprossessing and semantic dictionary mapping by showing free-text reports and the corresponding processed texts side-by-side. In our corpus, average word count of original free-text reports is 285 and the average word count of processed reports is 98, which is approximately 3x reduction in size.

\subsection{Word and Report Embedding}

After pre-processing and dictionary mapping, the corpus of 10,000 processed reports (see examples in Figure~\ref{fig:preprocessing}) was used to create vector embeddings for words in a completely unsupervised manner using the word2vec model that can be trained on a large text corpus to produce dense word vectors. Two unsupervised algorithms were introduced to obtain word to vector representation: Continuous Bag of Words (CBOW) and Skip-gram \cite{mikolov2013efficient}. Those algorithms learn word representations that maximize the probabilities of a word given other contextual words (CBOW) and of a word occurring in the context of a target word (Skip-gram).

Our semantic dictionary mapping step considerably reduced the size of our vocabulary by mapping the words in corpus to their root terms, thereby making the words in the vocabulary more frequent. CBOW is several times faster to train than the Skip-gram, with slightly better accuracy for frequent words. The CBOW architecture also captures the semantic regularities of words. Thus, CBOW approach appeared to be more suitable to be integrated into our framework, and, as expected, results of preliminary experiments with Skip-gram and CBOW showed CBOW to be the better performing model. 


We first constructed a vocabulary from our pre-processed tokenized corpus that contains 10,000 free-text radiology reports, and then learned \textit{vector representations of words} in the vocabulary. We build our predictive model using the Gensim 2.1.0 library~\cite{rehurek_lrec}. The CBOW word2vec model predicts a word given a context where context is defined by the window size. The loss function of CBOW is: $E = - {v_{w_{o}}}'.h + \log\sum_{j=1}^{V} \exp({v_{w_{j}}}'.h)$ , where $w_{o}$ is the output word, ${v_{w_{o}}}'$ is its output vector, $h$ is the average of vectors of the context words, and V is the entire vocabulary. Once the model constructs the vectors, we can use the cosine distance of vectors to denote similarity, thereby deriving analogies. The resulting word vectors can be used as features in many natural language processing and machine learning applications. 

As the training algorithm, we used both Hierarchical Softmax as well as Negative Sampling. Based on preliminary results, we found Negative Sampling to be better training algorithm. Mikolov et al. \cite{mikolov2013distributed} also described Negative Sampling as the method that results in faster training and better vector representations for frequent words, compared to more complex hierarchical softmax. The cost function of Negative Sampling is: $E = - \log  \sigma ({v_{w_{o}}}'.h) - \sum_{w_{j} \in \omega_{neg}} \log \sigma (-{v_{w_{j}}}'.h)$ , where $\omega_{neg}$ is the set of negative samples, $w_{o}$ is the output word, ${v_{w_{o}}}'$ is its output vector and h is the average of vectors of the context words.

Finally, the \textit{document vectors} were created by simply averaging the word vectors created through the trained model. According to Kenter et al. \cite{kenter2016siamese}, averaging the embeddings of words in a sentence has proven to be a successful and efficient way of obtaining sentence embeddings. Each document vector was computed as: $v_{doc} = \frac{1}{\left \| V_{doc} \right \|}\sum_{w \in V_{doc}}v_{w}$, where $V_{doc}$ is the set of words in the report and $v_{w}$ refers to the word vector of word $w$.

\subsection{Visualization of the embeddings \label{sec:WordEm}}
Our idea is to visualize the vector representation of words and documents to validate the semantic quality of the embeddings in two different levels. In the first level, the visualization of the trained individual word embeddings can verify the positioning of synonyms (and related words), antonyms and other word-to-word relations, and can show at the very low scale that if our vector embedding is able to preserve legitimate semantics of the natural words and clinical terms. Second, the visualization of the document vectors can fulfill the purpose of analyzing the proximity of documents that have different levels of likelihood of intracranial hemorrhage. If the documents corresponding to the same class (risk) appear close to each other and form clusters, we can infer that our embedding can be useful to boost the performance of any standard classifier. 
 
Our trained embeddings are expected to be high dimensional and may lie near a low-dimensional, non-linear manifold. Therefore, standard linear dimensionality reduction techniques (e.g. Principal Component Analysis) are not well-suited for preserving the distance between similar data points in low-dimensional representation of the vector space. We adopted t-Distributed Stochastic Neighbor Embedding (t-SNE) technique \cite{maaten2008visualizing} to visualize the trained embeddings using sklearn python library. t-SNE is a technique for dimensionality reduction that is particularly well suited to serve our application since it is capable of capturing much of the local structure of the high-dimensional data very well, while also revealing global structure such as the presence of clusters at several scales. It employs Gaussian kernel in the high-dimensional space and defines a soft border between the local and global structure of the data. For pairs of data points that are close to each other relative to the standard deviation of the Gaussian, t-SNE determines the local neighborhood size for each data point separately based on the local density of the data. We describe the results of t-SNE visualization of word and document vectors in the following section (Sec.~\ref{sec:VecVis}).

\subsection{Classification \label{sec:classify}}

In this study, the resulting document vectors were used as features to develop a computerized hemorrhage likelihood assessment system that aims to assign a \textit{`risk'} label to the free-text radiology reports while being trained on the subset of reports with the ground truth labels created by the experts (see Sec.~\ref{Sec:dataset}). We observed that our dataset had imbalanced distribution of training data, i.e. class 2, 3, and 4 had fewer instances than class 1 and 5. Thus, we grouped classes 2-4, and re-defined the class labels to ensure variation of the likelihood of intracranial hemorrhage as:  (1) `no risk' - no intracranial hemorrhage; 2) `medium risk' - probability of having intracranial hemorrhage; (3) `high risk'- definite diagnosis of intracranial hemorrhage. The re-definition of the class labels were validated by forming a mutual agreement between the two expert radiologists. In Table.~\ref{tab:no_exam}, we show the number of examples per class for the final three categories. To quantify the performance of the classifier, the 1,188 annotated reports were randomly divided into $80\%$ training set (950 reports) and $20\%$ test set (238 reports). To demonstrate the true power of our vector embedding, we performed experiments using three classifiers - Random Forests, Support Vector Machines, K-Nearest Neighbors~(KNN) in their default configurations. 

\begin{table}[]
\centering
\caption{Number of examples per category in our dataset}
\label{tab:no_exam}
\begin{tabular}{|c|c|l|l|}
\hline
\multicolumn{1}{|l|}{} & \multicolumn{3}{c|}{\textbf{Class labels}} \\ \hline
\multirow{2}{*}{\textbf{No. of cases}} & \multicolumn{1}{l|}{\textbf{`norisk'}} & \textbf{`medium risk'} & \textbf{`high risk'} \\ \cline{2-4} 
 & 946 & \multicolumn{1}{c|}{43} & \multicolumn{1}{c|}{199} \\ \hline
\end{tabular}
\end{table}

\subsection{Evaluation}
  We experimented with different types of kernels in SVM classifier (Radial kernel \& Polynomial kernel), and different values of `k' in kNN (k= 5,10) classifiers. To investigate the benefits of the proposed hybrid framework, we also tested each classifier's performance by creating vector embeddings of the radiology reports without the domain-specific semantic mapping (Sec.~\ref{sec:dicmapping}) where we skipped replacing the radiology terms and their synonyms using RadLex. However, we still substituted the common terms using the CLEVER base terminology for preserving the semantic structure of the radiology reports. In the Result section (see Sec.~\ref{sec:ClassiferPer}), we describe the performance of each classifier on the hold-out test set (238 reports) in a tabular format. Standard precision, recall and F1 score were used as metrics to quantify the classification performance.

\section{Results \label{sec:result}}

\subsection{Word analogies}
On feeding the entire corpus to the system, the final size of the resulting vocabulary was 4,442 words. We created word embeddings or semantic vector representations of words appearing in the corpus, from which several kinds of analogies could be derived by computing the similarity.  The similarity score between the word vectors was computed as cosine similarity which is inner product on the normalized space that measures the cosine of the angle between two words: $Similarity= \frac{A\cdot B}{\left \| A \right \| \left \| B \right \|}= \frac{\sum_{i = 1}^{n}A_{i}B_{i}}{\sqrt{\sum_{i = 1}^{n}A_{i}^{2}}\sqrt{\sum_{i = 1}^{n}B_{i}^{2}}}$. Table~\ref{table:synonyms} shows some synonyms/closely associated words and the cosine similarity scores of their respective word embeddings. Table~\ref{table:antonyms} shows some antonyms and the cosine similarity scores of their respective word embeddings. The data demonstrate that the system has formed embeddings such that pairs of synonyms have high similarity scores while antonyms have negative similarity scores.

\begin{table}[h]
\centering
\caption{Similarity scores of word embeddings of synonyms/closely associated words \label{table:synonyms}}
  \begin{tabular}{|l|l|c|}
  \hline
    \textbf{Word 1}    & \textbf{Word 2}  & \textbf{Similarity} \\
    \hline
    new & recent & 0.941 \\ 
    \hline
    overinflated & balloon\_appears & 0.999 \\ 
    \hline
    infarction & evidence\_hemorrhagic\_conversion & 0.910 \\ 
    \hline
    infarction & acute\_infarction & 0.928 \\ 
    \hline
    hemorrhage & rightward\_midline\_shift & 0.958 \\ 
    \hline
    hemorrhage & subdural\_hemorrhage & 0.964   \\ 
    \hline
    hemorrhage & intraventricular\_hemorrhage & 0.959 \\ 
    \hline
    hemorrhage & subarachnoid\_hemorrhage & 0.968 \\ 
    \hline
  \end{tabular}
\end{table}

\begin{table}[h]
\centering
\caption{Similarity scores of word embeddings of antonyms, NEGEX represents negation and QUAL represents severe terms (see Sec.~\ref{sec:dicmapping})   \label{table:antonyms}}
  \begin{tabular}{|l|l|c|}
  \hline
    \textbf{Word 1}    & \textbf{Word 2}  & \textbf{Similarity} \\ \hline
    large & NEGEX\_enlarged & -0.245   \\ 
    \hline
    hemorrhage & NEGEX\_QUAL\_hemorrhage & -0.074 \\ 
    \hline
    hemorrhage & NEGEX\_QUAL\_intracranial\_hemorrhage & -0.245 \\ \hline
    infarction & NEGEX\_QUAL\_infarction & -0.070 \\ 
    \hline
    large\_territory\_infarction & NEGEX\_QUAL\_large\_territory\_infarction & -0.157 \\ 
    \hline
    midline\_shift & NEGEX\_QUAL\_midline\_shift & -0.206 \\ 
    \hline
    abnormalities & NEGEX\_QUAL\_abnormalities & -0.283 \\ 
    \hline
    mass\_effect & NEGEX\_QUAL\_mass\_effect & -0.170 \\ 
    \hline
  \end{tabular}
\end{table}

\subsection{Vector Visualization \label{sec:VecVis}}
\begin{figure}[H]
\centering
\includegraphics[width=\linewidth]{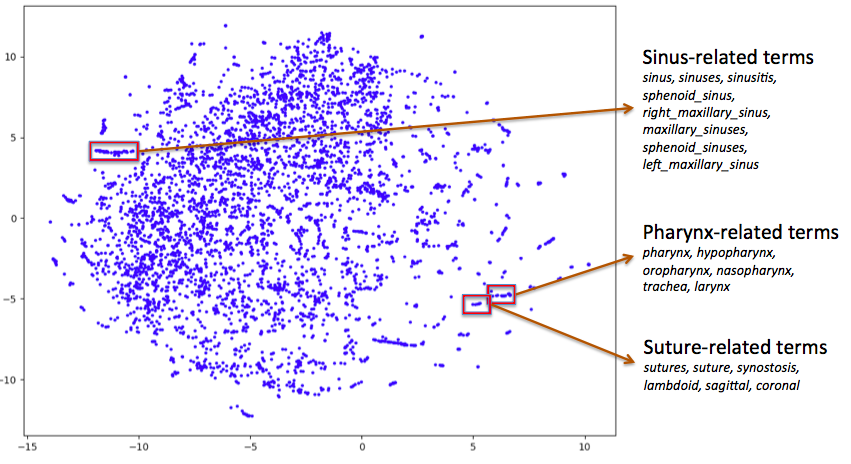}
\caption{All word embeddings (4,442 words) - visualized in two dimensions using t-SNE}
\label{fig:WordEmbedding}
\end{figure}

Figure~\ref{fig:WordEmbedding} shows the 2D visualization of word vector embedding constructed using the t-SNE approach (Sec.~\ref{sec:WordEm}) where each data point represents a word. A total of 4,442 words are visualized in the figure. As seen from the figure, similar words reside fairly close together and form a cluster in the map without even inclusion of any prior knowledge. This map illustrates that our word embedding can preserve semantics of the terms.          

\begin{figure}[h!]
\centering
\includegraphics[scale=0.4]{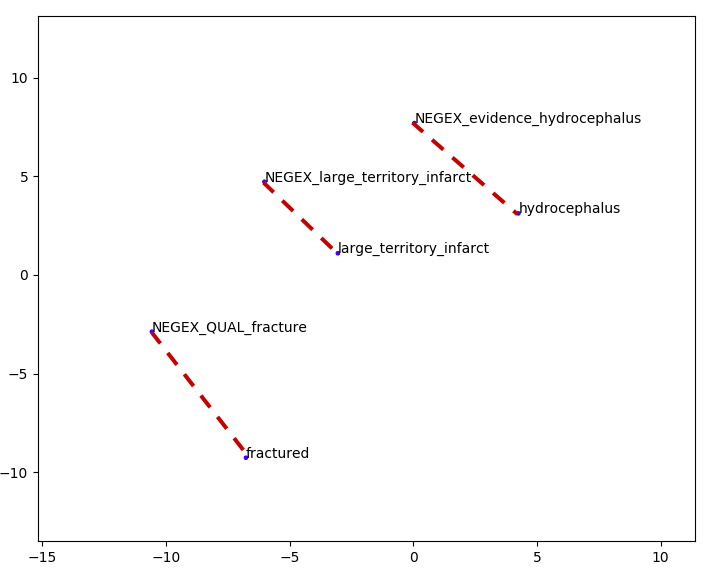}
\caption{Word embeddings: relation between terms and their negation}
\label{fig:WordNegation}
\end{figure}

In Figure~\ref{fig:WordNegation}, we also highlight a group of clinical terms particularly relevant for this case-study and their negations using the same t-SNE visualization technique. The figure illustrates ability of the embedding to automatically organize concepts and implicitly learn the relationships between them. To show the word-to-word relations, we visualize only a few significant terms and their negations, but same technique can be used to infer other analogies among the terms present in our vocabulary (e.g. synonyms, antonyms, finding-finding, finding-diagnosis).

\begin{figure}[h!]
\centering
\includegraphics[scale=0.45]{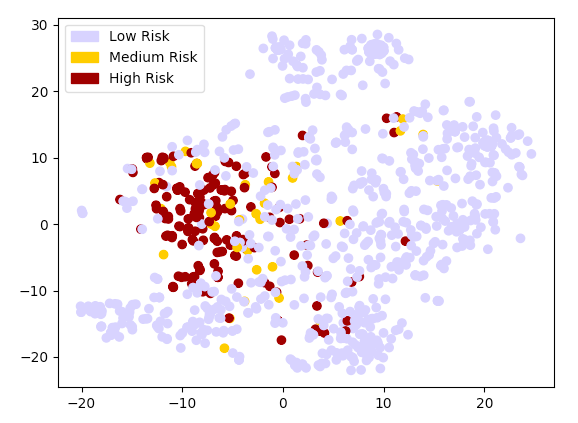}
\caption{1,188 CT Head radiology report vectors visualized in two dimensions}
\label{fig:DocEmbedding}
\end{figure}

We also visualize the subsequent vectors of complete reports projected in two dimensions using the t-SNE technique (Figure~\ref{fig:DocEmbedding}). This visualization has been created only for the 1,188 annotated reports since the main idea is to see if our proposed embedding can be useful to compute clusters with varying risk factors. From the Figure~\ref{fig:DocEmbedding}, we can see that the reports denoting high risk of intracranial hemorrhage cluster together, and the reports with intermediate risk are mostly residing close to high risk reports. Though this is a two dimensional projection of the original high dimensional document vector, the result clearly shows that the embeddings carry signals that could be very informative to automatically annotate the reports using state-of-the-art classifiers.  

\subsection{Classification performance \label{sec:ClassiferPer}}

We used the document vectors to classify each report into one of three classes denoting varying likelihood of intracranial hemorrhage (see Sec.~\ref{sec:classify}). As mentioned earlier in the paper, our radiology report embedding is flexible enough to be combined with both parametric and non-parametric classifiers. We experimented with three state-of-the-art classifiers - Random Forests, Support Vector Machines and K-Nearest Neighbors (KNN).     

\begin{figure}[H]
\centering
\includegraphics[scale=0.23]{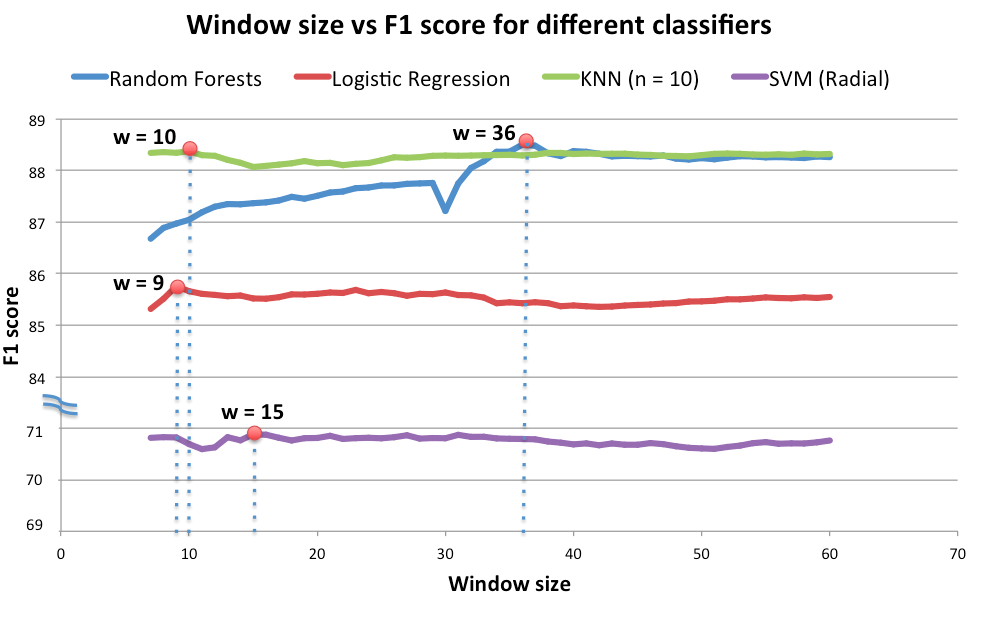}
\includegraphics[scale=0.23]{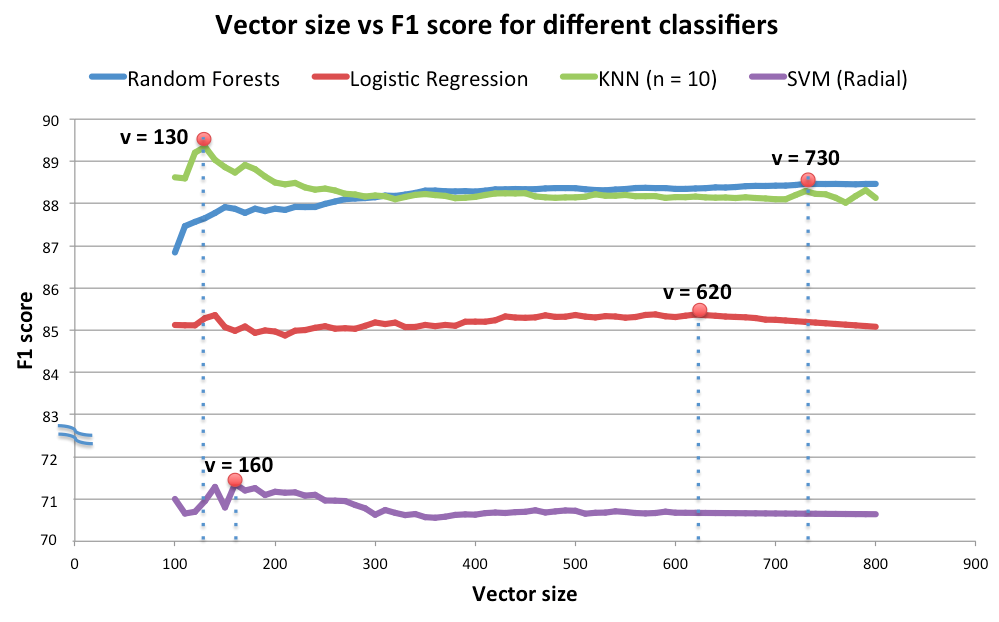}
\caption{Hyperparameter optimization of the embeddings using grid search: window size on the left and vector dimension on the right}
\label{fig:hyperparameters}
\end{figure}

To give more insight into the quality of the learned vectors, we used the grid search approach to tune the two main hyperparameters of our embedding for the targeted annotation, i.e. \textit{Window Size} and \textit{Vector Dimension}. The hyperparameter search was done individually for each classifier using cross-validation on the training data set. The effects of the hyperparameters on the resulting classifier performance are shown in Figure \ref{fig:hyperparameters} where the optimal points of the classifier's performance are highlighted. Based on the optimal points, we selected the hyperparameters and evaluated the classifiers' performance on the test set. For instance, Random Forest was evaluated with the word embeddings that were created with window size 36 and vector dimension 730. The standard classifiers are intentionally applied in their default configurations (as in the scikit learn framework) to demonstrate the ability to achieve high performance using the embedding created by our pipeline and improvement of performance over unigrams and out-of-the-box word2vec. \\
 
The classifiers' performance on the test set is reported in Table~\ref{table:performance} with optimal hyperparameters. We also present performance of the classifiers only using unigrams as features which can be considered as the baseline performance to be compared with word embedding. While the reported performance accuracy (F1-score) of baseline with unigrams is on average $71\%$, the word embedding resulted F1 score over $80\%$ for most cases which demonstrates that our vector representation was able to capture the significant facets of the radiology reports. The Random Forest classifier yielded a weighted precision of 88.64\% and weighted recall of 90.42\% with 730 dimensional word vectors, and closely outperforms all the other classifiers used in this study. However, KNN~$(n = 10)$ produces a weighted precision of 88.60\% and weighted recall of 89.91\% that is close to the Random Forest's performance, employing a reduced optimal word vector dimension~(130).

In Table \ref{table:performance}, we present the classifiers' performance with and without dictionary mapping as well as with unigrams as feature. In general, the word embedding improves the performance of the baseline classifiers and every classifier's performance is consistently better with the proposed hybrid technique. However, performance difference is incremental for the particular case study which is hypothesized to be due to the choice of dataset in which all the reports are associated to a very narrow domain and from the same institution, i.e. CT Head reports, and thus the variation in the vocabulary is relatively small. We expect that superiority in the performance of the proposed hybrid method may be more significant when multi-topic and multi-institutional free-text reports will be considered where the semantic and syntactic variations are more prominent.

\begin{table}[]
\centering
\caption{Performance of different classifiers with and without semantic mapping, and with unigrams features.}
\label{table:performance}
\resizebox{\textwidth}{!}{%
\begin{tabular}{|c|c|c|c|c|c|c|l|l|l|}
\hline
\textbf{} & \multicolumn{3}{c|}{\textbf{With Domain-specific dictionary}} & \multicolumn{3}{l|}{\textbf{Without Domain-specific dictionary}} & \multicolumn{3}{l|}{\textbf{Baseline with unigrams feature}} \\ \hline
\textbf{Classifier} & \textbf{Precision} & \textbf{Recall} & \textbf{F1 score} & \multicolumn{1}{l|}{\textbf{Precision}} & \multicolumn{1}{l|}{\textbf{Recall}} & \multicolumn{1}{l|}{\textbf{F1 score}} & \textbf{Precision} & \textbf{Recall} & \textbf{F1 score} \\ \hline
\textbf{Random Forests} & 88.64\% & 90.42\% & 89.08\% & 87.59\% & 89.17\% & 87.78\% & 87.5\% & 66.03\% & 75.26\% \\ \hline
\textbf{KNN (n = 10)} & 88.60\% & 89.91\% & 88.88\% & 86.73\% & 88.90\% & 87.47\% & 64.79\% & 80.49\% & 71.8\% \\ \hline
\textbf{KNN (n = 5)} & 88.54\% & 89.62\% & 88.76\% & 87.52\% & 88.65\% & 87.74\% & 82.62\% & 82.36\% & 75.9\% \\ \hline
\textbf{SVM (Radial kernel)} & 64.19\% & 80.09\% & 71.25\% & 63.98\% & 79.96\% & 71.07\% & 60.52\% & 77.80\% & 68.08\% \\ \hline
\textbf{SVM (Polynomial kernel)} & 63.25\% & 79.49\% & 70.43\% & 62.40\% & 78.97\% & 69.70\% & 60.52\% & 77.80\% & 68.08\% \\ \hline
\end{tabular}%
}
\end{table}

\section{Conclusion \label{sec:conclusion}}
In this study, we have shown how to efficiently learn dense vector representations of individual words as well as entire radiology reports by using a hybrid technique that combines word2vec and semantic dictionary mapping. Our experimental results show that our proposed embeddings were able to learn the actual semantics of the radiological terms from free-text reports. Thanks to the embeddings, we successfully annotated the radiology reports according to the likelihood of intracranial hemorrhage with 89.08\% F1 score. We have publicly released (\href{https://github.com/imonban/RadiologyReportEmbedding} {https://github.com/imonban/RadiologyReportEmbedding})  our trained embeddings that have been used to test the classifiers performance (Table~\ref{table:performance}), which can be directly reused to support similar radiological applications, e.g. inferring relations between clinical terms, annotation of radiology reports, etc. The techniques introduced in this paper can be used also for creating vector representation from clinical notes of different domains (e.g. oncology) given a domain-specific ontology that can be used to reduce underlying term variations in the corpus. 

In the prospective future studies, we will compare alternative neural word embedding methods (e.g. GloVe) since we believe that the performance of any such method  will be boosted by the semantic mapping, as the models are initialize with random vector for out-of-vocabulary words which is far for reality. In the future version of the pipeline, we will incorporate log-likelihood ratio and mutual information for identify frequently appearing pairs, and will consider different linear functions (max pool, average pool, min pool etc.) to create document embedding from word vectors. 

\section*{Acknowledgement}
This work was supported in part by grants from the National Cancer Institute, National Institutes of Health, U01CA142555, 1U01CA190214, and 1U01CA187947

\centering
\bibliographystyle{unsrt}

\end{document}